\documentclass[onecolumn]{els-mrw_modified} 

\usepackage{amsmath,amssymb,amsfonts,amsthm,makeidx,graphicx}
\usepackage{txfonts}
\usepackage{helvet}
\usepackage{braket}
\usepackage{hyperref}

\def\be{\begin{equation}}
\def\ee{\end{equation}}
\def\bea{\begin{eqnarray}}
\def\eea{\end{eqnarray}}

\def\bi{\begin{itemize}}
\def\ei{\end{itemize}}
\usepackage{dsfont}

\usepackage{tikz}
\definecolor{lime}{HTML}{A6CE39}
\DeclareRobustCommand{\orcidicon}{%
    \begin{tikzpicture}
    \draw[lime, fill=lime] (0,0) 
    circle [radius=0.13] 
    node[white] {{\fontfamily{qag}\selectfont\tiny ID}};
    \draw[white, fill=white] (-0.0625,0.095) 
    circle [radius=0.007];
    \end{tikzpicture}
    \hspace{-2mm}}
   
\newcommand{\orcidJK}{\href{https://orcid.org/0000-0003-0998-9460}{\orcidicon}}

\begin{document}

\chapter{Many-body localization}\label{chap1}

\author[1,2]{Jakub Zakrzewski\orcidJK}%

\address[1]{\orgname{Uniwersytet Jagielloński}, \orgdiv{Instytut Fizyki Teoretycznej,  Wydział Fizyki, Astronomii i Informatyki Stosowanej}, \orgaddress{ulica Stanisława \L{}ojasiewicza 11, PL-30-348 Krak\'ow, Poland}}
\address[2]{\orgname{Jagiellonian University}, \orgdiv{Mark Kac Complex   Systems Research Center}, \orgaddress{PL-30-348 Krak\'ow,
    Poland}}

\articletag{Chapter Article tagline: Sept. 19, 2025}

\maketitle

\begin{glossary}[Keywords]
Nonergodic Quantum Dynamics, Many-Body Physics, Thermalization

\end{glossary}

\begin{abstract}[Abstract]
We present an introductory review of nonergodic dynamics in interacting many-body quantum systems, focusing on the phenomenon of many-body localization (MBL). We describe aspects of MBL and summarize the evidence for a crossover from the ergodic to the MBL regime in finite systems, using the paradigmatic XXZ model as an example. We then broaden the scope to other models to illustrate the generality of the phenomenon. We briefly touch on the largely unexplored relation between MBL and quantum computing.
\end{abstract}

\section{Introduction}\label{chap1:sec1}

A standard definition of “quantum chaos” is the behavior of quantum systems that are classically chaotic in the semiclassical limit. This traditionally involves few-body nonintegrable situations where such a limit is well controlled by, loosely speaking, taking all relevant quantum numbers to infinity (this is not necessarily the case for conserved quantum numbers; an example is the hydrogen atom in a constant magnetic field, where a semiclassical limit can be considered within a single orbital quantum number, e.g., the $L_z=0$ sector).

For many-body physics the situation is somewhat different. Except for a few idealized problems, most models involve interacting fermions or spins that remain intrinsically quantum. Instead of a semiclassical limit, one typically considers the thermodynamic limit in which the system size (or number of particles) tends to infinity. 

The dynamics of such systems are typically thermalizing. For models coupled to an environment this is well defined: a subsystem prepared in an arbitrary initial state and in contact with a large reservoir thermalizes, and its long-time state can be described by an appropriate Gibbs ensemble. The subsystem therefore loses memory of its initial state. By contrast, for a finite isolated system the whole evolution is unitary and time-reversible, so in principle the initial state can be recovered. This apparent contradiction underlies the eigenstate thermalization hypothesis (ETH), which provides our current understanding of generic quantum many-body dynamics. Dynamics that obey ETH are loosely termed ergodic; their basic properties are discussed in Section~\ref{sec:ETH}.

As with every “generic” phenomenon, there are exceptions. One class is formed by fully integrable, disorder-free quantum models; these lie beyond the scope of this chapter. Instead, we consider a robust ergodicity-breaking phenomenon: many-body localization (MBL), typically associated with sufficiently strong disorder. MBL is also often regarded as stemming from an effective integrability for individual disorder realizations (see Section~\ref{sec:MBL}). We discuss in some detail the measures and tools used to characterize MBL and the challenges associated with its thermodynamic limit. We also address the interesting case of recently identified positional disorder, for which standard local MBL diagnostics appear to fail. In subsequent sections, we discuss disorder-free localization arising from Hilbert-space fragmentation as well as other interacting, experimentally relevant models that may lead to MBL. Finally, we briefly consider the implications of possible quantum computing for the progress in our understanding of localization in many-body systems. Due to space (and editor) limitations, we provide a selected list of references. While attempting to cite many important works, we apologize for omissions and refer readers to several reviews on MBL \cite{Nandkishore15, Alet18, Abanin19, Gopalakrishnan20, Sierant25} for more extensive bibliographies.   

\section{Eigenstate thermalization hypothesis}\label{sec:ETH}


The introduction below is elementary. For more insight see {\color{red} Quantum Thermalization} chapter \cite{Patil26}.

Consider an isolated many-body system described by a Hamiltonian $\hat H$ with eigenstates $|m\rangle$ and eigenvalues $E_m$. The time evolution
of an observable $\hat{A}$ for an initial state $|\psi\rangle=\sum c_n |n\rangle$ 
implies that the average value of $\hat{A}$ after time $t$ reads
\begin{equation}
 A(t)  = \sum_{m} |c_m|^2 A_{mm} + \sum_{m,n\neq m}
 \mathrm{e}^{-\mathrm{i} (E_m-E_n) t} c^*_m c_n A_{mn},
 \label{eq:A1}
\end{equation}
where $A_{mn} = \langle m |\hat A | n \rangle$. The time average of $A(t)$ over time $T$ is given by
\begin{equation}
 \bar{A}(T) = \frac{1}{T} \int_0^{T}dt A(t) \stackrel{T \to \infty }{\longrightarrow} \sum_m |c_m|^2 A_{mm}.
  \label{eq:A2}
\end{equation}
The long-time behavior of $A$ (and the possibility of thermalization) is determined by the diagonal matrix elements $A_{mm}$ (we assume a non-degenerate spectrum for simplicity).
The time scale for relaxation to equilibrium is controlled by the off-diagonal elements $A_{mn}$ and by the energy differences $\omega_{mn} = E_m-E_n$ (compare Eq.~\eqref{eq:A1}) that enter the expansion of the initial state in the energy eigenbasis.

The standard formulation of ETH yields an ansatz for both diagonal and off-diagonal matrix elements of observables in the eigenbasis of $\hat H$ \cite{Srednicki99} (for reviews see \cite{Polkovnikov11, Dalessio16}): 
\begin{equation}
 A_{mn} = \mathcal{A}(\bar E) \delta_{mn} + \mathrm{e}^{-S(\bar E)/2}f_{\mathcal A} (\bar E, \omega_{mn}) R_{mn},
 \label{eq:ETH1}
\end{equation}
where  $\bar E = (E_m+E_n)/2$ is the mean energy 
and $R_{mn}$ is a random variable
with zero mean and unit variance.
The function $S(\bar E)$ is the thermodynamic entropy at energy $\bar E$ (the logarithm of the density of states), and $\mathcal{A}(\bar E)$ and $f_{\mathcal A} (\bar E, \omega)$ are smooth functions. By contrast, in random matrix theory ({\bf RMT}) one has $\mathcal{A}(\bar E) = {\rm const}$ and $f_{\mathcal A} (\bar E, \omega) = 1$. In ETH, $\mathcal{A}(\bar E)$
is the expectation value in the microcanonical ensemble while $f_{\mathcal A} (\bar E, \omega)$ 
carries information on the nature of transport and the fine structure of response functions \cite{Dalessio16}. The former may be seen by
expanding \eqref{eq:ETH1} for the diagonal matrix elements around the mean energy $\braket{E}=\braket{\psi|\hat{H}|\psi}$, obtaining 
\cite{Rigol08} 
\begin{equation}
 \lim_{T\to \infty}\bar{A}(T) = \frac{1}{\mathcal N_{\langle E \rangle, \delta E}}
 \sum_{|E_m - \langle E \rangle| < \delta E} A_{mm} \equiv \mathrm{Tr} \left[ \hat{\rho}_{MC} \hat{A} \right],
 \label{eq:ETH2}
\end{equation}
where $ \hat{\rho}_{MC}$ is the microcanonical density matrix and $\mathcal N_{\langle E \rangle, \delta E}$ is the number of states in the energy window $\delta E$ around the mean energy $\braket{E}$. 
Therefore, the long-time average of the observable $\hat{A}$ is independent of the specific values of the coefficients $|c_m|^2$, and equals the average of $A_{mm}$ over the entire energy shell $|E_m - \langle E \rangle| < \delta E$, i.e., the microcanonical average. 
ETH is thus believed to provide a sufficient criterion for the onset of thermalization~\cite{Dalessio16}.

ETH implies general statistical properties of the Hamiltonian matrix. Its eigenvalues follow RMT predictions, exhibiting level repulsion with a distribution of normalized spacings (nearest-neighbor eigenvalue differences $s_n=E_{n+1}-E_n$) well approximated by the Wigner distribution corresponding to the appropriate symmetry class of random matrices (orthogonal for time-reversal-invariant systems and unitary otherwise \cite{Haakebook})\footnote{The third universality class, the symplectic ensemble, is quite special and rarely encountered; see \cite{Haakebook} for details.}
Spacings must be normalized (unfolded) because they typically vary with energy due to the changing density of states; it is therefore useful to consider the dimensionless ratio of consecutive spacings, defined as $r_n=\mathrm{min}(s_n/s_{n+1}, s_{n+1}/s_n)$ \cite{Oganesyan07}. While exact distributions are known (under the Wigner surmise) \cite{Atas13}, we note that the average gap ratio for the GOE (Gaussian orthogonal ensemble) is $\overline r_{\rm GOE}=0.5307$, for the GUE (Gaussian unitary ensemble) it is $\overline r_{\rm GUE}=0.5996$, and for a Poisson level sequence—corresponding to an integrable system or a superposition of independent spectra - $\overline r_{\rm P}=0.3863$. 

Properties of eigenstates in ETH systems may be characterized by their entanglement entropies. If we divide the system into a region A and its complement B, the expectation value of a local observable $\hat A$ (supported on A) for an eigenstate $|n\rangle$ should be thermal. Hence the reduced density matrix $\rho_A=\mathrm{Tr_B} |n\rangle \langle n |$ is thermal, implying that the entanglement entropy $S=-\mathrm{Tr} \rho_A \log \rho_A$ is extensive and grows with system size—it follows a “volume law”.   

It is worth mentioning that systems obeying ETH typically also exhibit diffusive transport. ETH is a hypothesis and there is no rigorous criterion delineating when it holds or fails. It is widely expected to apply to physical observables in generic systems, but several examples violate ETH. Apart from strictly integrable systems, the most robust example is many-body localization.


\section{Many-body localization}
\label{sec:MBL}
Studies of MBL commenced seriously with independent efforts \cite{Gornyi05,Basko06} that indicated, using a perturbative approach, that for sufficiently large disorder, transport in interacting one-dimensional (1D) systems may be blocked indicating localization. That triggered intensive research over the last twenty years, full of often diverse claims. The subject has been reviewed several times \cite{Nandkishore15, Alet18, Abanin19,Sierant25}. The most recent review \cite{Sierant25} concentrates on attempts to determine the MBL/ergodic transition in the thermodynamic limit as postulated by an early work \cite{Pal10}. Let us state immediately that this problem, at present, remains open. Referring the interested reader to the review \cite{Sierant25}, we shall rather concentrate on finite systems that may, at best, undergo a crossover from an extended, delocalized, ergodic regime to localized behavior manifestly breaking the ETH. Our approach may be considered pragmatic, since any experiment, even in the most isolated ultracold-atom setting, is performed with a finite number of particles as well as for a finite time.

\subsection{Spectral measures}

\begin{figure}
 \includegraphics[width=0.45\linewidth]{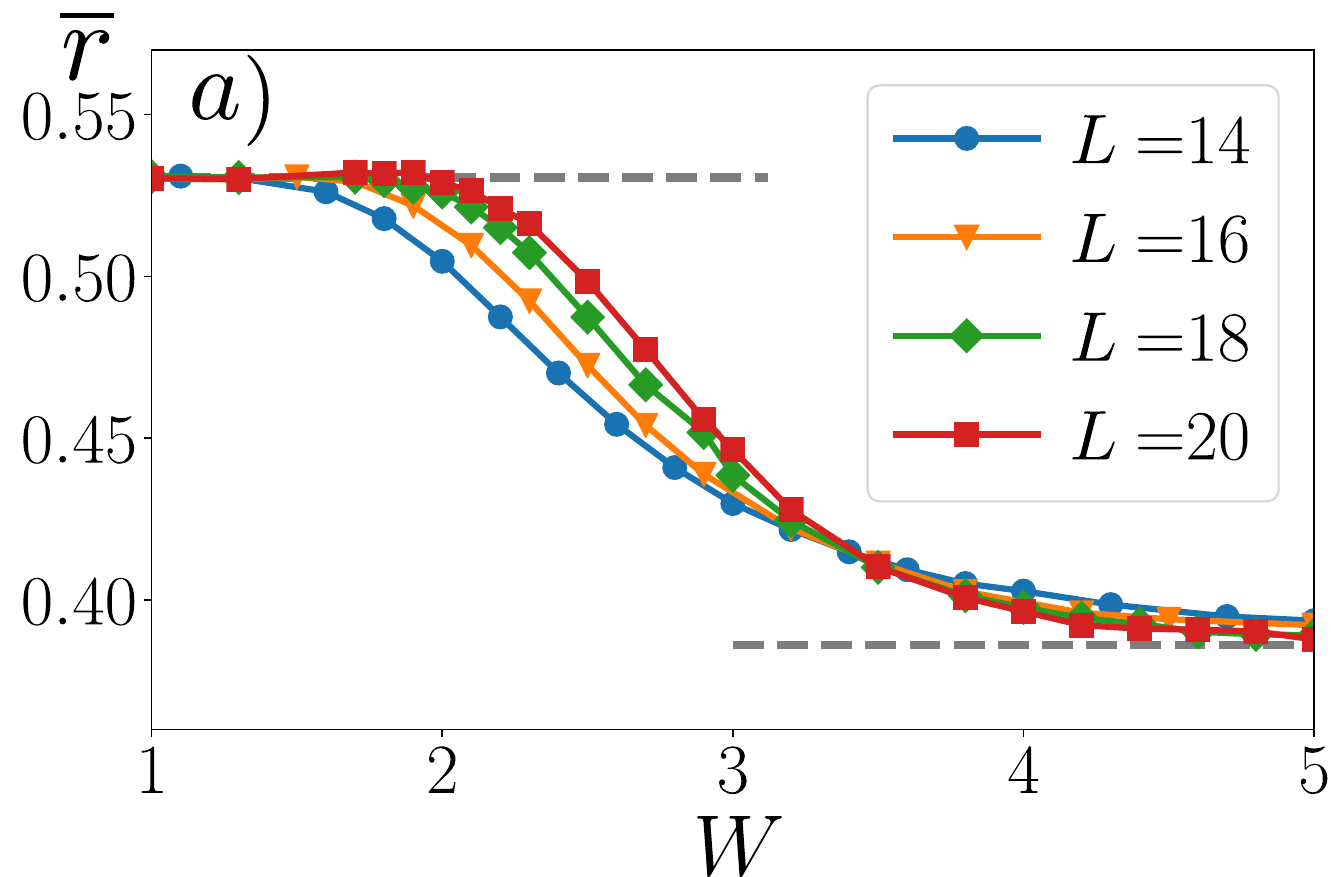}
 \includegraphics[width=0.45\linewidth]{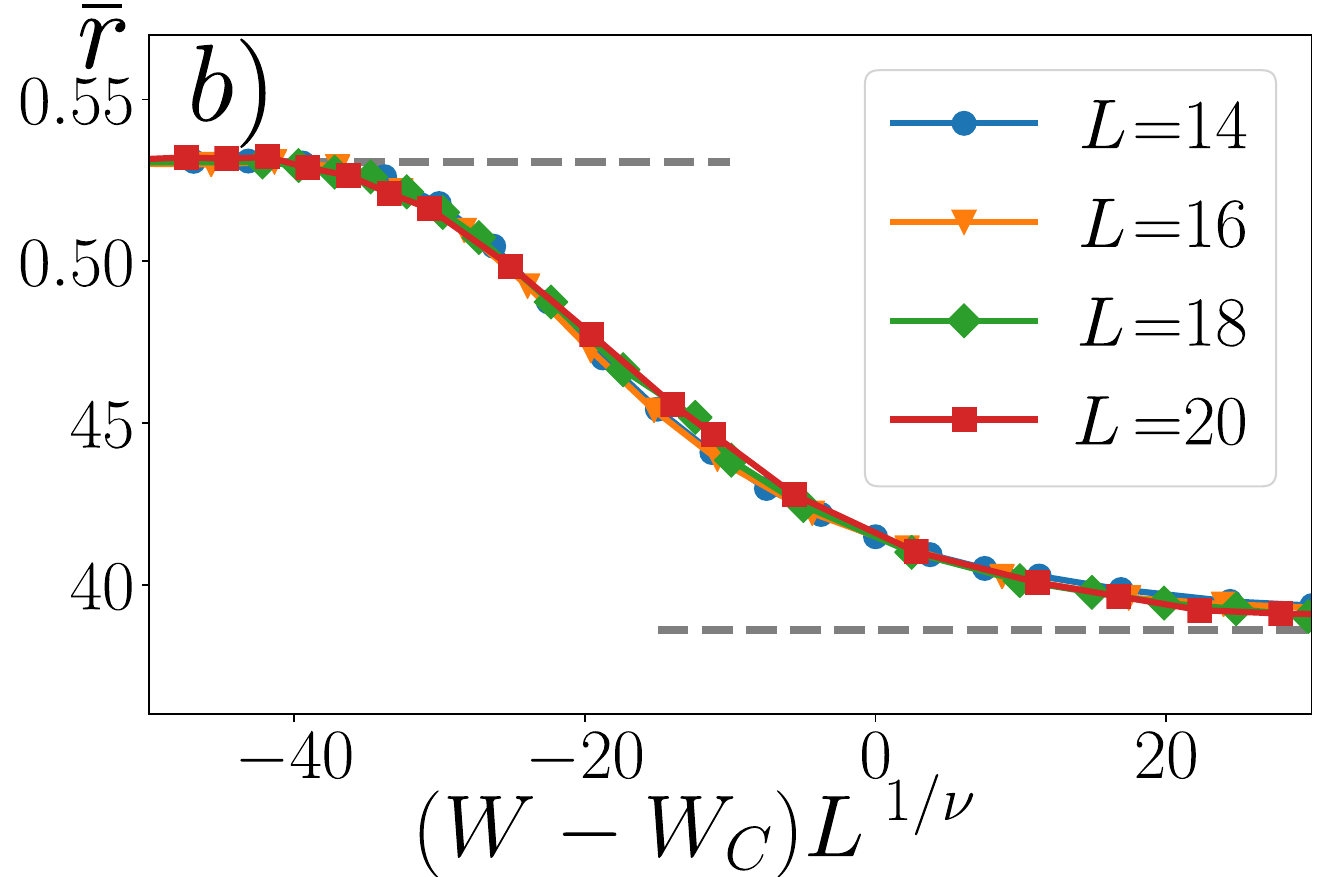}
  \caption{ {(a) Average gap ratio as a function of the disorder amplitude shows the crossover from GOE to Poisson values, indicating the crossover from ergodic to localized regimes. (b) Finite-size scaling yields the putative critical disorder amplitude $W_c\approx3.7$; see, however, the comments in the text. 
  \label{fig:gap} 
 }}
\end{figure}

Among the possible interacting one-dimensional models, the disordered Heisenberg chain has achieved the status of the paradigmatic case of studies. The Hamiltonian reads:
\begin{equation}
 H= J\sum_{i=1}^{L} \ \vec{S}_i \cdot \vec{S}_{i+1} + \sum_{i=1}^{L} h_i S^z_i,
 \label{eq: XXZ}
\end{equation}
where  $\vec{S}_i$ are spin-1/2 matrices, $J=1$ is fixed as the energy unit, periodic boundary conditions are assumed $\vec{S}_{L+1} = \vec{S}_1$, and $h_i \in [-W, W]$ are independent, uniformly distributed random variables. This model shows a clear crossover from ergodic (for small $W$) to localized regimes as revealed by gap-ratio statistics—cf. Fig.~\ref{fig:gap}.

\begin{BoxTypeA}[chap1:box1]{Standard MBL properties}

Many-body localization is a property of (at least) a part of the excited spectrum of a many-body interacting and disordered system characterized by the following correlated characteristics:
\begin{itemize}
\item the existence of a complete set of local integrals of motion (LIOMs), separate for each disorder realization, implying quantum integrability;
\item statistical properties of spectra corresponding to the Poisson ensemble;
\item area-law entangled eigenstates;
\item a long-lasting memory of the initial nonstationary state in the time dynamics; 
\item a logarithmic growth of the entanglement entropy in a quench from an initially low-entanglement state. 
\end{itemize}
Note that the above is assumed to hold for a part of the excited spectrum. There are systems that show a mobility edge that separates localized and ergodic regimes in energy.
\section*{Possible extensions}
The above properties are observed, typically, for quasi-local Hamiltonians with diagonal (i.e., on-site) disorder. Systems showing strongly nonergodic and often localized behavior also occur when, for some reason, the corresponding Hilbert space is strongly fragmented (shattered). Nonergodic features occur often for systems with high global symmetries. Those symmetries, necessarily obeyed by eigenvectors, lead to at least sub-volume entanglement as they must respect the global symmetry.
\end{BoxTypeA}

The crossings of curves for different system sizes [panel (a)] enable finite-size scaling (b), after which both the critical disorder value, $W_c=3.7$, and the correlation critical exponent $\nu=0.9$ are extracted. 
The latter is unexpected; for a critical transition, according to the Harris bound, one needs $\nu d>2$ (where $d=1$ is the system dimension). Clearly, this bound is violated. A similar analysis may be carried out for entanglement entropy, where the change from the volume-law behavior (proportional to the chain length $L$) to the area-law (constant) is observed by looking at $S/L$ scaling or the rescaled entropy, $s_E=S/S_P$, where $S_P$ is a typical value for random matrices, referred to as Page value \cite{Page93average}, the peak in entropy fluctuations, or related measures \cite{Luitz15}.

The breaking of Harris's bound suggests, however, that the finite-size scaling approach may be misleading. 
Already in the early study \cite{Oganesyan07}, significant shifts with the system size of the crossings in gap ratio were observed, pointing out the difficulty in determining the existence of an MBL phase based on spectral statistics alone [compare also Fig.~\ref{fig:gap}(a)].

The very existence of the MBL transition was questioned on estimations based on spectral form-factor analyses \cite{Suntajs20e} and the drift of the disorder value, $W_T$, at which $\bar r$ differs from the GOE value, with the system size: $W_T\sim L$. Spectral form factor allows one to estimate the Thouless time $t_{\rm Th}$—the time after which the dynamics of the system is universal and faithful to the appropriate (concerning symmetries) random-matrix ensemble. It was demonstrated \cite{Suntajs20e} that the Thouless time scales as 
 $   t_{\rm Th}= t_0 e^{W/\Omega}L^2,$
where constants $t_0, \Omega$ depend on the microscopic details of the model. The polynomial increase of $t_{\rm Th}$ with the system size should be confronted with the exponential scaling of the Heisenberg time (which represents the timescale after which the discreteness of the spectrum manifests itself). Thus, for sufficiently large $L$, the Thouless time must become smaller than the Heisenberg time, implying ergodic behavior. It was later shown \cite{Sierant20thouless}, however, that multidimensional Anderson models, known to have a localized phase in the thermodynamic limit, share a similar scaling of the Thouless time for small sizes. This scaling clearly breaks for large systems for the three-dimensional Anderson transition. The breakdown of the scaling could be demonstrated in the random XXZ model, where the largest system sizes available to state-of-the-art numerics violate it.

\begin{figure}
 \includegraphics[width=0.9\linewidth]{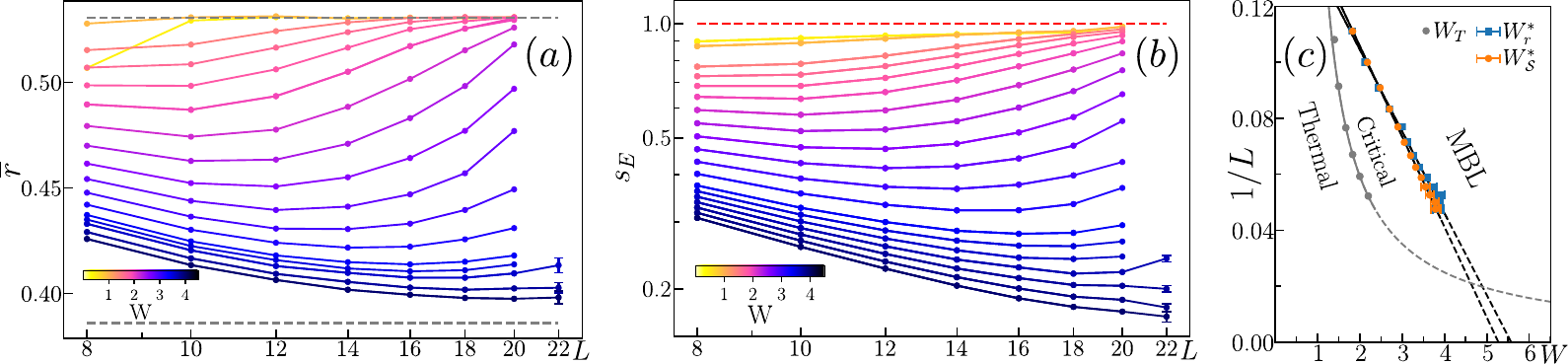}
  \caption{ Average gap ratio $\bar r$ (a) or the rescaled entropy $s_E$ (b) as a function of the system size, $L$, for different, color-coded disorder amplitudes, enabling the extraction of crossover disorder values $W_T$, $W_{r}^*$, $W_{S}^*$ plotted in panel (c) for different system sizes. The extrapolating dashed lines intersect near $W\approx 5$ at $L\approx 50$; see the text for discussion. 
  \label{fig:sca} 
 }
\end{figure}

While the arguments of \cite{Suntajs20} were also system-size affected \cite{Sierant20p}, one can understand the problem by inspecting Fig.\ref{fig:sca}. Observe that both $\bar r$ and $s_E$ curves for sufficiently large disorder values show minima as a function of $L$ turning upwards (towards ergodic values) for too large $L$.
These minima may serve as estimates for the MBL border.
Extracting the positions of the minima $W_{r}^*(L)$, $W_{S}^*(L)$ shows that their drift with $L$ is $\sim 1/L$, as shown in Fig.\ref{fig:sca}(c). For system sizes reachable by exact diagonalization, non-zero critical regions separate the low-disorder, thermal (ETH) regime from the localized (MBL) one. Since exact diagonalization does not presently allow reaching larger system sizes, one may extrapolate both borders as shown by dashed lines. They cross around $W=5$ for $L=50$. The extrapolation of the MBL border to $L\rightarrow\infty$ yields an estimate $W_c\approx 5.5$.  

Similar conclusions may be reached by studying properties of eigenvectors. Here, the crossover to the MBL regime may be observed by, e.g., the mean entanglement entropy behavior. In the ETH regime, it follows the typical Page value \cite{Page93average}, growing practically linearly with the one-dimensional system size (i.e., revealing the so-called volume-law scaling). In the localized regime, the eigenstates exhibit area-law entanglement, which is practically independent of the system size (i.e., volume); thus, the ratio of the mean entanglement entropy to the Page value decays rapidly to zero. Again, finite-size scaling of the mean-entropy curves can be carried out (see, e.g., \cite{Khemani17, Alet18}). The detailed analysis of the so-called participation entropies \cite{Mace19Multifractal} has shown that eigenstates on the “localized” side are multifractal. That may be linked to a visualization of MBL as Anderson localization in Fock space, as reviewed in \cite{Roy24}. Among multiple other studies of various properties, we mention a detailed analysis of spin–spin correlation properties 
\cite{Colbois25} that points toward the existence of system-wide instabilities even in the “deep” MBL regime. Such system-wide resonances were earlier indicated to exist in the MBL regime, shifting the border of “true” MBL to large disorder values \cite{Morningstar22}.


Other authors \cite{Sels21,Sels23}, however, question the very existence of MBL in the thermodynamic limit by examining the sensitivity of eigenstates to adiabatic transformations as addressed by fidelity susceptibility and the low-frequency asymptotes of the spectral function. Fidelity susceptibility, commonly used for ground states to identify quantum phase transitions \cite{Zanardi06}, can be defined for any eigenstate $|n\rangle$ by analyzing its changes with a change of some relevant parameter of the Hamiltonian $\lambda$ as
\begin{equation}
    \chi_n=\langle n | \overleftarrow{\partial}\lambda \partial\lambda |n\rangle = \sum_{m\ne n}\frac{|\langle n|\partial_\lambda \hat H|m\rangle|^2}{(E_n-E_m)^2}.
\end{equation}
Interestingly, exact analytic distributions of $\chi_n$ for models faithful to Gaussian ensembles were found \cite{Sierant19f}.
As with gap ratios, it is highly informative to inspect not the full distributions but the mean fidelity susceptibility or the typical log-fidelity susceptibility defined as $\xi= \langle \log(\chi_n)\rangle$, where the mean is taken over different eigenstates and different disorder realizations. Then the typical scaled fidelity susceptibility, $\exp(\xi)/2^L$, shows a clear peak for the XXZ model that drifts to larger disorder values with increasing system sizes. Authors \cite{Sels21} argue that this indicates exponentially long relaxation times. By accompanying this with an analysis of the low-frequency dependence of the spectral form factor, which points toward subdiffusive, very slow behavior for sufficiently large disorder values $W>1$, the authors conclude that in the thermodynamic limit localization is destabilized due to a strong divergence of the matrix elements of operators at small energy differences at the level of the smallest level spacings, dominating the tail of the distribution of $\xi_n=\log(\chi_n)$.

\subsection{Quasi-local integrals of motion}

The apparent crossover to localized behavior with Poissonian level-spacing statistics suggested that in the MBL regime (at least for finite systems) there exists a complete set of quasi-local integrals of motion (LIOMs) \cite{Serbyn13b,Huse14}. The generic Hamiltonian for such a model may be expressed as:
\begin{equation}
    \hat{{H}}_{\mathrm{MBL}}= \sum_{i} h_i\hat{\tau}_i^{z} + \sum_{i<j}J_{ij}\hat{\tau}_i^{z}\hat{\tau}_j^{z} + \sum_{i<j<k}J_{ijk}\hat{\tau}_i^{z}\hat{\tau}_j^{z}\hat{\tau}_k^{z} + ...
    \label{eq:l-bit}
\end{equation}
where $h_i$ are random on-site fields, $J_{ij\ldots}$ are interaction terms that decay exponentially with the distance between the spins, and $\hat{\tau}_i^{z}$ are quasilocal operators that mutually commute — the LIOMs. They may be roughly viewed as spin $\hat{S}_i^{z}$ operators dressed with exponentially decaying weights by nearby $\hat{S}_{i-1}^{z},\hat{S}_{i+1}^{z},\hat{S}_{i-2}^{z},\hat{S}_{i+2}^{z},\dots$
This picture allows one to draw several conclusions about the dynamics \cite{Serbyn14}. In particular, for a global quench, local observables reach stationary, nonthermal values (dependent on the initial state) at long times due to the slow dephasing typical of the MBL phase. Thus, owing to the existence of LIOMs, the system exhibits some form of memory.
Temporal fluctuations around stationary values exhibit universal power-law decay in time, with an exponent set by the localization length. This is related to the logarithmic in time growth of entanglement in the MBL phase for initially low-entanglement states (e.g., separable), as observed numerically \cite{Znidaric08,Bardarson12}.
These predictions have been partially tested in experiments as described below. Note that the existence of a full set of LIOMs, rendering the system integrable, leads to vanishing transport. Thus, subdiffusive dynamics may serve as a signature of integrability breaking.

\subsection{Time dynamics}

\subsubsection{Time-correlation function — imbalance}
While spectral studies address at most 26 interacting spins, experiments can treat much larger systems, as exemplified by the Munich group experiment \cite{Schreiber15}. The information is obtained by studying the time dynamics of nonstationary states. In the experiment, interacting fermions were prepared in an optical lattice, with odd sites filled by single fermions and even sites empty. The observable followed was the imbalance between occupations of odd and even sites ($N_o$, $N_e$, respectively), defined as $I(t)\equiv (N_o-N_e)/(N_o+N_e)$. The ETH dynamics would lead to an equalization of the occupations of lattice sites and a vanishing imbalance. Disorder was introduced by adding secondary lattice beams that shifted the lattice minima in a quasiperiodic fashion as $h_i=W\cos(\beta i +\phi)$, where $\beta$ is an irrational number, $i$ the site index, and $\phi$ a random variable changing from realization to realization. For sufficiently large $W$, the imbalance saturated at a nonzero value during the experimental time, indicating that the system remembered its initial state, thus exhibiting MBL.

Similarly, the imbalance may be studied in spin chains. One can define (more formally) the imbalance as the disorder-averaged (denoted by the overline) two-time spin correlation function $C(t)$ averaged over sites:
\begin{equation}
 I(t) = \overline{C(t)} = D \sum_{i=1+l_0}^{L-l_0} \overline{ \langle \psi(t) |\hat S^z_i | \psi(t)\rangle \langle \psi | \hat S^z_i |\psi \rangle}
 = D \sum_{i=1+l_0}^{L-l_0} (-1)^i \overline{ \langle \psi(t) |\hat S^z_i | \psi(t)\rangle},
 \label{eq:imb}
\end{equation}
where $| \psi(t)\rangle = e^{-i H t} |\psi\rangle$, $\ket{\psi}$ is the initial state, and the constant $D$ ensures that $I(0)=1$. The parameter $l_0>0$ diminishes boundary effects. The third equality result assumes the initial state to be the N'eel state with every second spin up and the others down, $\ket{\psi} = | \uparrow \downarrow \ldots \uparrow \downarrow\rangle$. Such a state lies roughly in the middle of the spectrum of the XXZ model, so it is typically taken as the initial state for the time dynamics. The latter may be studied exactly for slightly larger system sizes than those accessible to exact diagonalization, thanks to Chebyshev propagation schemes, and for large system sizes by tensor-network approaches (see \cite{Sierant22challenges} for details and further references).

\begin{figure}
 \includegraphics[width=0.9\linewidth]{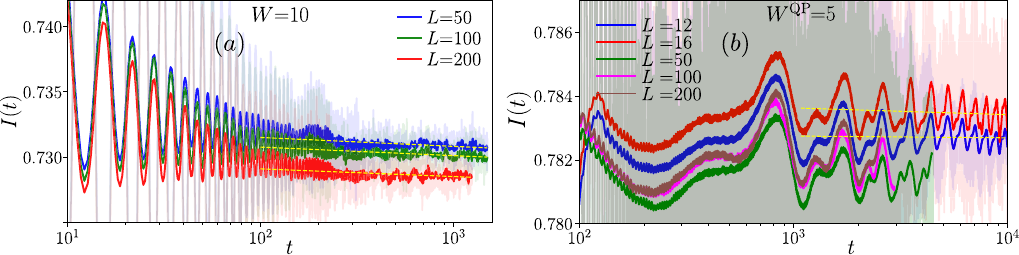}
  \caption{ Time evolution of imbalance I(t) for the Heisenberg spin chain for different system sizes as indicated. Left panel corresponds to random uniform disorder with $W=10$, right panel to QP disorder with $W^{QP}=5$ as indicated in the panels. Shaded lines give I(t), while solid lines denote a running average of I(t) over the window $(t-25, t+25)$; dashed yellow lines denote power-law fits $I(t)\sim 1/t^\beta$ in the time interval $t \in [100, 1500]$ with $\beta \approx (3.5\pm 0.5)\times 10^{-4}$ for the random disorder. Modified from P. Sierant, J. Zakrzewski (2022) \cite{Sierant22challenges}.
  \label{fig:imb} 
 }
\end{figure}
The obtained $I(t)$ curves reveal, for large disorder values, a slow decay in time which, on some time intervals, may be considered a power-law decay, $I(t)\sim 1/t^\beta$. Small-system-size studies show that the decay persists beyond the Heisenberg time, which grows exponentially with the system size, $L$, as $t_H\sim \exp(aL)$. It follows that at $t_H$, $I(t_H)\sim \exp(-a\beta L)$, i.e., it vanishes in the $L\to\infty$ limit. Therefore, does MBL survive in this limit? It may happen that $\beta$ decays with system size, or that it slowly decreases with time, allowing a nonzero imbalance in the long-time limit. As shown in \cite{Sierant22challenges}, in the numerically accessible range, $\beta$ does not depend on system size and appears constant for large times (limited in TN simulations) even for disorder values $W>10$ — see Fig.~\ref{fig:imb}; for a detailed discussion, see \cite{Sierant25}. A plausible physical explanation of these numerical results is via the existence of system-wide many-body resonances due to very broadly distributed effective matrix elements in strongly interacting disordered chains \cite{Morningstar22}, which shift the border of MBL to around $W\sim 20$ for the Heisenberg chain. Those resonances manifest as strongly correlated spin pairs traversing the entire system (cat states with system-wide correlations) \cite{Laflorencie25}. 
They may be responsible for a partial breaking of LIOMs leading to a broad distribution of their relaxation times \cite{Vidmar21phenomenology}. 
Some authors cast doubt on the existence of a true MBL phase in the thermodynamic limit \cite{Sels21}.

\subsubsection{Entanglement entropy}

Dynamics of other observables, not as easily accessible experimentally as the imbalance, have also been studied, yielding similar controversies. As mentioned above, the LIOMs approach indicates that the entanglement entropy growth for an initially separable (or, more generally, weakly entangled) state grows logarithmically in time \cite{Serbyn13a}, as already shown numerically in \cite{Znidaric08}; see also \cite{Bardarson12}.

The disordered XXZ spin chain conserves the $z$-component of the total spin, $\hat{S}^z \equiv \sum_{i=1}^{L} \hat{S}^z_i$, i.e., the $U(1)$ symmetry (corresponding to particle-number conservation in the fermionic language). The density matrix of the subsystem $A$ splits into orthogonal sectors, $\hat{\rho}_A = \sum_s p_s \hat{\rho}_{A,s}$, enabling an analysis of symmetry-resolved entanglement, in particular the splitting of the entanglement entropy between subsystems $A$ and its complement $B$ into the configurational entanglement entropy, $S_c$, and the number entropy, $S_n$ \cite{Schuch04, Schuch04b}. Here, $p_s$ is the probability of finding the system in sector $s$ (in the studied case, corresponding to the possible eigenvalues of total $S_z$ in subsystem $A$, i.e., a given number of particles in $A$), while $\hat{\rho}_{A,s}$ is the density matrix of subsystem $A$ in sector $s$. The entanglement entropies are
\begin{equation}
S = S_n + S_c,\quad \text{with}\quad
S_n = -\sum_s p_s \ln p_s,\quad
S_c = -\sum_s p_s \operatorname{Tr}_A\bigl(\hat{\rho}_{A,s}\ln(\hat{\rho}_{A,s})\bigr).
\label{eq:number}
\end{equation}
The number entropy, $S_n$, describes the distribution of the number of particles (partial $\hat{S}_z$ projections) among sectors, while the configurational entanglement entropy, $S_c$, is related to different spin configurations appearing within sectors.

For an interacting delocalized many-body system, the number entropy is expected to grow logarithmically: $S_n(t) \propto \ln t$ \cite{Kiefer20}. In the MBL phase, one expects that particle-number fluctuations between the subsystems saturate at long times \cite{Bardarson12}, leading to the saturation of $S_n(t)$. Instead, \cite{Kiefer20} observed numerically a slow double logarithmic increase of the number entropy, $S_n(t) \propto \ln\ln t$. This was confirmed in a TDVP study \cite{Sierant22challenges} for $L=50$, deep in the MBL regime of the disordered XXZ spin chain at $W=10$. The observed growth of $S_n(t)$ was interpreted \cite{Kiefer20} as arising from remaining slow particle transport, implying the absence of MBL. Instead, at finite times one observes \cite{Kiefer21, Kiefer21Unlimited} $S(t) \propto \ln t$ and $S_n(t) \propto \ln\ln t$.

This conclusion was questioned first by \cite{Luitz20}, where the double-logarithmic growth of $S_n(t)$ was interpreted as a transient feature related to single-particle motion across the boundary between subsystems. Subsequently, \cite{Ghosh22} associated the growth of the number entropy with rare resonances, pointing out that the double-logarithmic growth observed may be described by another functional form, leading to saturation at long times. The heated discussion followed \cite{Kiefer22comment, Ghosh22response}. Interestingly, it has been shown that the double-logarithmic growth of $S_n(t)$ in time is also obtained within an artificial LIOM model \cite{Chavez2023ultraslow}, which is obviously localized.

\subsubsection{Internal clock of many-body dynamics}
\label{intri}

An interesting observation was made in \cite{Evers23Internal}: the entanglement entropy, $S(t)$, may serve as an intrinsic artificial time for the evolution. Different disorder realizations yield different, e.g., spin-correlation functions $C(t)$ (compare with \eqref{eq:imb}). Instead of looking at $C(t)$ and its average over disorder, $I(t)$, one can consider $C(S)$ and proceed to average over disorder the correlation functions at a given $S$. If indeed it is the entanglement entropy that determines the imbalance value, the proposed quantity $\overline{C(S)}$ should exhibit much smaller sample-to-sample fluctuations than $I(t)$. In the ergodic regime, entanglement spreads ballistically, with $S(t)\propto t$, and the fictitious time measured by $S(t)$ is proportional to real time.

As shown in \cite{Evers23Internal}, the imbalance and its variance, plotted as functions of the disorder-averaged entanglement entropy $\overline{S(t)}$, collapse onto a single curve depending solely on the system size for different disorder values $W<8$ in the XXZ spin chain. Thus, $\overline{S(t)}$ is indeed an internal time measure (clock), enabling the identification of universality in the regime of slow dynamics. Therefore, for the considered system sizes $L\le 24$ and time scales, there are no visible signatures of the MBL phase. For any disorder strength $W$, the decay of the imbalance is dictated by the growth of entanglement entropy, $\overline{S(t)}$. An MBL phase would require the emergence of another branch of the curve for $L\to\infty$, along which the fictitious time grows (since $S(t)\propto \ln t$) and the autocorrelation functions do not decay. The absence of such a branch for times $t<10^3$ considered in \cite{Evers23Internal} is consistent with the persistent decay of the imbalance $I(t)$ shown in Fig.~\ref{fig:imb}. Therefore, if an MBL phase exists, the evidence for it should consider larger time scales and system sizes.

\subsection{Two dimensional models}

We have discussed one-dimensional physics up to this point. For two (2D) or more dimensions, studies are scarcer. The seminal experiment \cite{Choi16} showed that in the presence of a sufficiently strong disorder, the part of the 2D system that is initially occupied remains dominantly populated even at long times, suggesting MBL. On the other hand, the experiment could be simulated with the Gutzwiller mean field approach \cite{Yan17} that cannot describe the entanglement inherent in MBL. Very recent studies  \cite{Hur25} suggest that in 2D for random disorder, the MBL crossover point shifts to higher disorder strength with increasing system size, consistent with the avalanche scenario. The latter  \cite{DeRoeck17} provides an analytic analysis of the avalanche growth of chaotic grains as resulting from small local disorders and negates the existence of MBL in more than one dimension. 

\section{Quasiperiodic disorder}

\begin{figure}
\centering
 \includegraphics[width=0.7\linewidth]{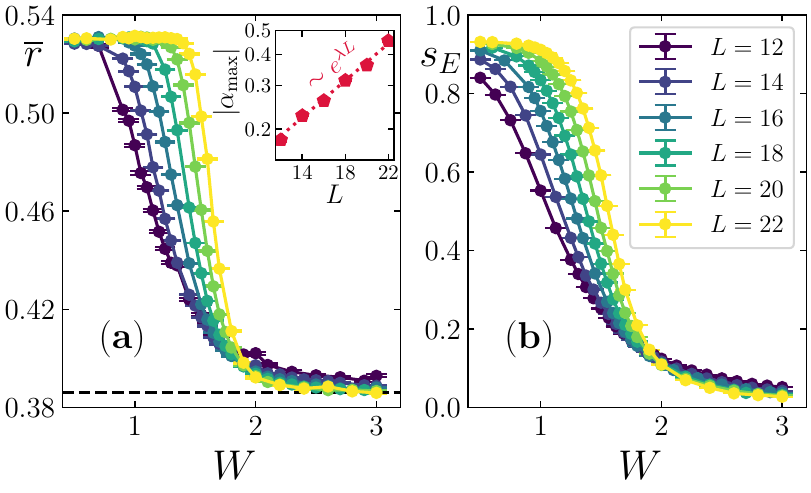}
   \caption{ { Mean gap ratio (a) and rescaled by the Page value mean entanglement entropy (b) as a function of the system size for XXZ model with QP on-site disorder. The insets show exponential scaling of the maximal slope, $|\alpha$, of the curves.  Modified from  Aramthottil et al. (2021) \cite{Aramthottil21}.
  \label{fig:slope} 
 }}
\end{figure}
As mentioned above, while most of the theoretical works assume a random uniform disorder, the experiments (see e.g. \cite{Schreiber15, Hur25}) favor an easy-to-realize with laser beams induced potentials, quasiperiodic (QP) disorder. The lack of rare random Griffiths regions leads to less pronounced 
size effects as discussed, e.g., in \cite{Aramthottil21}. While the border of the ETH behaviour shifts linearly with the system size $W_T\sim L$ for the random uniform disorder, for the QP case, one observes $W_T\sim \ln L$, still, however, diverging with the system size. Moreover, by calculating the derivative of $\overline r(W)$ or  $\overline S(W)$ curves, one finds that their maximal modulus of the slope grows exponentially with $L$, indicating an increased sharpness \cite{Falcao24}. By comparison, for a random disorder, the slope increases linearly only. Similar behavior is also observed for other models of disordered interacting systems. 

Note that by lowering the primary optical lattice potential, one may realize other QP effective disorders \cite{Kohlert19}. For such a shallow lattice, one has to take into account tunneling to the next nearest neighbors. The combination of different laser wavelength beams results in the so-called generalized Aubry-Andre\'e model that reveals the so-called mobility edge even for a noninteracting case (see \cite{Luschen18} for the experimental realization). In such a case, one naively expects lack of localization for the interacting model, which is, however, not the case.  The signature of nonergodic MBL-like dynamics of imbalance can also be observed for this model \cite{Kohlert19}.

The lack of rare Griffiths regions of relatively small disorder in the quasiperiodic disorder suggests that localization in more than one dimension may be possible \cite{Agrawal22}. A recent 2D experiment \cite{Hur25} finds a weak dependence on the system size, suggesting the existence of MBL. Still, as discussed by us above, even in one dimension, the fate of MBL phase in the thermodynamic limit in quasiperiodic disorder is not fully clear. 
\section{Avalanche scenario and the Quantum Sun}
\label{sec:qs}

The problem with establishing the existence of MBL in the thermodynamic limit is linked with the avalanche scenario \cite{DeRoeck17}, 
which describes how a small chaotic grain may expand in a many-body interacting system. 
This scenario rules out the existence of MBL in more than one dimension, 
while in one-dimensional chains it supports the existence of MBL for interactions that decay exponentially with distance. 
On the other hand, in purely random disorder, rare Griffiths regions may appear at arbitrary sizes, effectively destabilizing MBL. 
That may explain to some extent the numerical system size dependence described above.

The same picture allows to construct a model with minimal or no size effects, the so-called Quantum Sun (QS) model \cite{Suntajs22}. 
It consists of a small chaotic grain called the sun (modeled by $N$ 1/2-spins enjoying all-to-all interactions) 
and sun-rays, i.e., couplings of $L\gg N$ 1/2-spins to the random sun. 
Those $L$ spins do not interact with each other. 
Each such spin is coupled to one randomly chosen spin of the sun. 
The $j$-th outer spin is at the distance $u_j$ from the sun, with $u_j$ chosen uniformly from the $[j-\zeta,j+\zeta]$ interval, 
while its coupling to the spin from the sun decays exponentially with the distance $g_0\alpha^{u_j}$.
The outer spins are also exposed to a random magnetic field chosen uniformly from $[0.5,1.5]$. 
Internal degrees of freedom of the sun are described by a $2^N\times 2^N$ random matrix 
from the Gaussian Orthogonal Ensemble (GOE). 
The values of parameters chosen in \cite{Suntajs22} are: $N=3, \zeta=0.2, g_0=1, \beta=0.3$.

The close link of this model with the avalanche picture may be realized by noting 
that the outer spins are nothing but approximate LIOMS (so they do not interact among themselves) 
while the chaotic grain tends to destabilize them. 
The exponential decay of interactions of the grain with outer spins correlates with the fact that LIOMS are quasi-local; 
their support decays exponentially with distance, and this is reflected by the couplings assumed. 
Depending on the range of these exponential interactions, controlled by $\alpha$, one observes an extended or localized character of the QS.

The original QS model does not conserve total spin projection on the z-axis $\sum S_j^z$ 
(i.e., does not conserve the particle number in the fermionic language).
Such a modified model, preserving $U(1)$ symmetry, may be realized \cite{Pawlik24} 
by restricting the GOE matrix to the appropriate symmetry sector and appropriately modifying the interactions. 
The Hamiltonians for both models read
\begin{equation}
\hat{H}=R\otimes\mathds{1}+\sum\limits_{j=1}^{L}g_0\alpha^{u_{j}}\hat{S}^{x}_{n_{j}} \hat{S}^{x}_{j}+\sum\limits_{j=1}^L h_j\hat{S}^z_j; \quad\quad
\hat{H}_{\mathrm{cons}}=R_{\mathrm{cons}}\otimes\mathds{1}+\sum\limits_{j=1}^{L}g_0\alpha^{u_{j}}(\hat{S}^{x}_{n_{j}} \hat{S}^{x}_{j}+\hat{S}^{y}_{n_{j}} \hat{S}^{y}_{j})+\sum\limits_{j=1}^L h_j\hat{S}^z_j,
\label{eq:sun}
\end{equation}
where $R\in \mathrm{GOE}$ and $R_{\mathrm{cons}}$ is its symmetry-constrained version.
The original model closely resembles, in fact, the avalanche theory idealized case \cite{DeRoeck17, Suntajs24}. 

\begin{figure} \label{fig:qsun} 
\centering
 \includegraphics[width=0.9\linewidth]{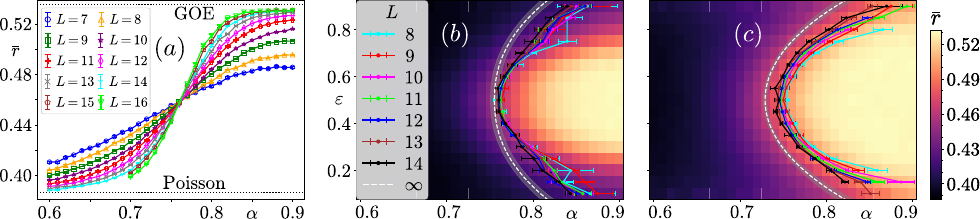} \quad
   \caption{ (a) Mean gap ratios $\overline r$ in the middle of the spectrum (as a function of
the coupling power $\alpha$ for chains with $L$ outside spins 
for the U(1) Hamiltonian) cross at a single $L$-independent point at $\alpha\approx 0.76$.
Dotted lines indicate the values expected from
ergodic (GOE) and integrable (Poisson) systems. Panels (b) and (c) show $\overline r$ for the full spectra, 
revealing the mobility edge even in the extrapolated (dotted white) $L\rightarrow\infty$ regime. 
Modified from Pawlik et al. (2024) \cite{Pawlik24}.
 }
\end{figure}

The model allows for a clear identification of the transition from the ergodic phase occurring for large $\alpha$ 
to the localized phase with the transition point close to $\alpha_c=1/\sqrt{2}$ as predicted by theory. 
This is revealed both by spectral form factor analysis and mean gap ratio \cite{Suntajs22, Pawlik24}. 
The latter is shown for the U(1) case in Fig.~\ref{fig:qsun}(a). 
Note the negligible drift with the system size of the crossing separating localized and extended regions. 
The data in panel (a) corresponds to the middle of the spectrum. 
The analysis of all eigenvalues shows that both the original (b) and the U(1) symmetric (c) models 
reveal the mobility edge in the thermodynamic limit \cite{Pawlik24}, 
contradicting the claim of the lack of it for many-body disordered systems \cite{DeRoeck16}.

The Quantum Sun model indicates that there exist many-body interacting systems that convincingly reveal MBL in the thermodynamic limit. 
The question, however, remains to what extent the construction presented above is engineered to reveal the MBL transition 
and how general it is for disordered interacting systems. 
From some point of view, the Quantum Sun may be considered a 0-dimensional model, 
thus contributing little to the discussion on one-dimensional spin chains.

 \section{MBL due to Hilbert space shattering} 

 Disorder was an essential ingredient of the models discussed so far. 
Recently, however, it has been shown that strong localization may be observed in tilted chains \cite{vanNieuwenburg19,Schulz19}, 
where the standard XXZ Hamiltonian is supplemented by the term $H_T=F\sum_i in_i$ with $F$ being the tilt amplitude. 
For a sufficiently large $F>2$, one observes a clear transition to Poissonian level statistics, 
provided a small inhomogeneity is added to the potential (either a small harmonic term \cite{Schulz19} or a very small disorder \cite{vanNieuwenburg19}). 
Similarly to standard MBL, one observes the logarithmic growth of entanglement entropy from an initially separable state \cite{Schulz19}. 
The necessity of breaking tilt uniformity to clearly observe MBL-like features is very interesting, 
making a pure uniform tilt quite special. 
As explained in \cite{Schulz19}, a pure tilt leads to the appearance of numerous degeneracies of many-particle states 
with the same center of mass (or dipole moment $d=\sum_i in_i$) for a noninteracting system. 
Then the evolution from a Fock-like state leads to a rapid initial growth of entanglement when such states couple due to exact degeneracies \cite{Yao21a}. 
Only after sufficient time for interactions to take place does the entanglement growth become weak again. 

 Tilted chain localization seems to be a special example of localization due to Hilbert space shattering, 
as discussed in detail in \cite{Khemani20,Sala20}. 
Such situations occur if even a few global conservation laws exist, 
such as the particle number (charge) and the center of mass (dipole) in the tilted lattice case. 
This leads to Hilbert space fragmentation into many disjoint sectors that do not mix in the dynamics—a phenomenon termed ``Hilbert space shattering.'' 
As noted in \cite{Khemani20,Sala20}, the dynamics constrained by charge and dipole conservation makes the system akin to fracton systems 
characterized by a limited mobility of fractons (see e.g., \cite{Pretko20,Nandkishore19} and references therein).

 Such disorder-free localization has soon been observed experimentally both in tilted optical lattices \cite{Scherg21} 
and quantum simulators \cite{Guo20,Morong21}. 
Interestingly, however, an exact proof exists that the pure tilt case cannot produce localization \cite{Kloss23} 
and leads to subdiffusive transport instead. 
The subdiffusive character of transport has been confirmed independently by others as well \cite{Zhang20,Nandy24}.

 Leaving pure tilt aside, with an additional harmonic \cite{Schulz19} or weak disorder \cite{vanNieuwenburg19} potential, 
the tilted XXZ chain exhibits signatures typical of strong disorder localization (at least for finite systems), 
forming a prime example of localization due to Hilbert space shattering. 
Another example is provided by a strongly interacting gas of hardcore dipolar bosons in a one-dimensional optical lattice \cite{Li21}. 
The hardcore limit may be obtained if on-site interactions are so strong that double and higher occupations of sites are energetically very costly. 
Such a model maps into a spin-1/2 chain with interactions decaying as a power law with the distance between sites. 
For permanent dipoles (as realized by molecules), $V/r_{ij}^3$ behavior is expected with $r_{ij}=a|i-j|$, where $a$ is the lattice constant. 
The Hamiltonian becomes $H=-t\sum_i (a^\dagger_ia_{i+1} +h.c.)+ \sum_{i\ne j} V/r_{ij}^3a^\dagger_ia_{i}a^\dagger_ja_{j}$. 
If dipolar interactions are neglected beyond the nearest neighbors, the effective model becomes an XXZ chain with $\Delta=V/a^3$. 
Large $V$ leads to Hilbert space fragmentation and the appearance of an emergent approximate constant of motion $\sum a^\dagger_ia_{i}a^\dagger_{i+1}a_{i+1}$, 
which fragments the Hilbert space. 
Within each sector, however, the motion is delocalized. 
Taking into account a full dipolar tail leads to additional emerging approximate constants $\sum_i a^\dagger_ia_{i}a^\dagger_{i+k}a_{i+k}$ for $k=2,3\dots$, 
resulting in Hilbert space shattering and very slow dynamics that, for realistic evolution times, appear to be localized for a sufficiently large $V$.

\section{ Non-standard MBL - positional disorder}

Until now, we have considered mostly the case of diagonal disorder, either random or quasiperiodic. The exception has been the Quantum Sun model considered in Section~\ref{sec:qs}, where a small disorder was also added to the positions of the outer spins. Here we consider another, related possibility; namely, we allow the couplings between sites (the tunneling amplitudes or interaction strengths) to be disordered. Since interactions typically depend on distance, one may realize such a disorder by randomizing the positions of the interacting bodies. This would be rather difficult in an optical lattice potential but can be realized experimentally by placing atoms in optical tweezers \cite{Browaeys20,Signoles21}. Here, atoms are typically immobile (as trapping potentials are deep), but when they are excited to Rydberg levels, they may interact via enhanced dipolar interactions, effectively realizing interacting spin models \cite{Browaeys20}. The resulting Hamiltonian can be written as:
\begin{equation}
    \hat{H} = \sum_{i \neq j}^{L} J_{i, j} \left(\hat{S}^x_i \hat{S}^x_j + \hat{S}^y_i \hat{S}^y_j \right) + \sum_{i \neq j}^{L} \Delta_{i, j} \hat{S}^z_i \hat{S}^z_j, \qquad \mathrm{with} \quad  J_{i,j} = \frac{J}{|r_i - r_j|^n}, \ \ \ \ \Delta_{i,j} = \frac{\Delta}{|r_i - r_j|^m}
    \label{eq:bond_disorder_hamiltonian}
\end{equation}
 where $r_i$ denotes the $i$-th atom's position, $J$ and $\Delta$ are distance-independent constants, while the exponents $n$ and $m$ are typically $3$ or $6$, depending on the interaction regime in which the Rydberg atoms are prepared (for details, see \cite{Browaeys20}). 

\begin{figure} \label{fig:rgx} 
\centering
 \includegraphics[width=0.75\linewidth]{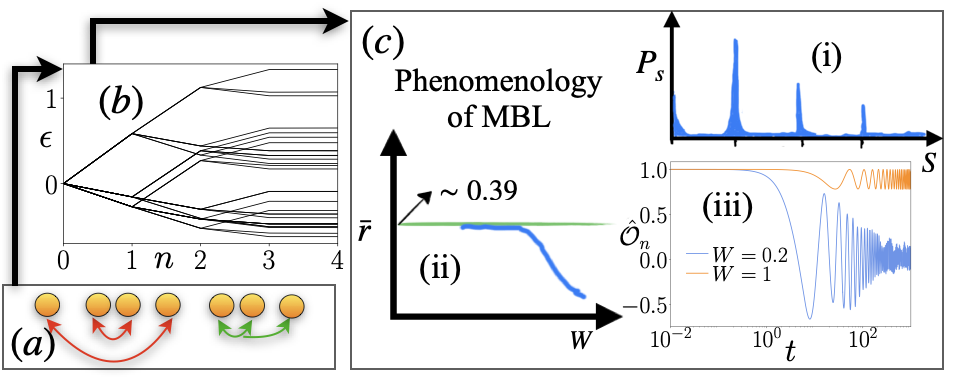}
   \caption{RSRG-X approximates eigenstates
and eigenvalues of the bond-disorder XXZ spin chain for large disorder. (a) RSRG-X finds and decimates strong bonds in the spin chain, yielding approximate eigenvalues, $\epsilon$, for a single realization of the XXZ Hamiltonian with M = 8 spins -- see panel (b).
(c) Properties captured by the RSRG-X procedure: (i) nontrivial entanglement entropy distribution, (ii) sub-Poissonian
mean gap ratio, (iii) lack of thermalization of $\mathcal{\hat O}_n$ at large $W$. Figure reprinted from Aramthottil et al. (2024) \cite{Aramthottil24}, copyright American Physical Society, 2024.}
\end{figure}

Consider $L$ atoms placed at random positions within a length $M$. The
Rydberg blockade prevents two atoms from being too close to each other. Thus, there is a minimal length for a system with equal site spacing—the Rydberg diameter. 
Increasing $M$ gives more freedom for site positions to become random, provided no two sites are closer than the Rydberg blockade allows. The larger the system size, the greater the randomness. One can then define an effective disorder in the system that is inversely proportional to atomic density.
As the disorder is not coupled to the on-site operators, the resulting localization is expected to differ significantly from the cases discussed in the preceding sections. In particular, it was demonstrated \cite{Aramthottil24} that the description of this limit in terms of LIOMs is no longer valid. On the other hand, one may use the phenomenological picture provided by the Real-Space Renormalization Group for excited states (RSRG-X) \cite{Pekker14}. This procedure addresses the strongest links first, diagonalizing the resulting small-scale local Hamiltonians, switching to the new basis, and iteratively expanding until the whole system is covered; it can be used to asymptotically reproduce the excited eigenstates of the model in the limit of sufficiently large disorder. This procedure is illustrated in Fig.~\ref{fig:rgx}(a), with the resulting typical level structure plotted in Fig.~\ref{fig:rgx}(b). Figure~\ref{fig:rgx}(c) illustrates the basic features of the bond-disordered model. It exhibits sub-Poissonian level spacing statistics for sufficiently large disorder, with the half-chain entanglement entropy distributions peaking around integer values (for the base-2 logarithm). This reflects an abundance of Schrödinger cat states often covering the whole system. Such pairs of cats are narrowly spaced, affecting the mean gap ratio value. Panel (iii) in Figure~\ref{fig:rgx}(c) shows the time dynamics of typical operators. The blue line shows the dynamics of a typical quasi-local operator, which thermalizes. On the other hand, one may build operators on two or more sites connected by strong bonds—such operators do not thermalize, as illustrated by the fate of $\hat{\mathcal{O}}_n$ (orange line).

\section{Beyond coupled 1/2-spins}

\subsection{Fermi-Hubbard chain and beyond}

Until now, we have discussed mostly chains of locally coupled 1/2-spins that, in a single dimension, map to hard-core bosons or, via Jordan-Wigner transformation, to spinless fermions.
On the other hand, some experiments have been carried out with unpolarized standard interacting fermions (see, e.g., \cite{Schreiber15, Kondov15}) as described by the disordered Fermi-Hubbard model. The Hamiltonian of the system reads:
\begin{equation}
\hat{H}
= - \sum_{\left<i, j \right>}^{L} \sum_{\sigma \in\{\uparrow, \downarrow\}} t_\sigma \hat{f}_{i, \sigma}^{\dagger}\hat{f}_{j, \sigma} 
 + U \sum_{i}^{L} \hat{n}_{i, \uparrow} \hat{n}_{i, \downarrow} + \sum_{i,\sigma}^{L} h_{i,\sigma} \hat{n}_{i, \sigma},
\label{eq:fh}
\end{equation}
where $\hat{f}_{i, \sigma}^{\dagger}$ ($\hat{f}_{j, \sigma}$) creates (annihilates) a fermion with spin $\sigma$ at site $i$, while $t_\sigma$ are tunneling coefficients (possibly different for different spins) and $\hat{n}_{i, \sigma}= \hat{f}_{i, \sigma}^{\dagger}\hat{f}_{i, \sigma}$. We assume an equal occupation of spin-up ($\uparrow$) and spin-down ($\downarrow$) fermions. If the tunnelings of both species are equal ($t_\downarrow = t_\uparrow$)
and the disorder is correlated (i.e., spin-independent), the system enjoys SU(2) symmetry. Then, on general grounds, one does not expect MBL \cite{Vasseur15, Potter16}. This can be simply explained by the fact that eigenstates, faithful to the global symmetries of the system, must possess information about the whole system. This is not possible for area-law entangled states.
In effect, spin degrees of freedom slowly delocalize \cite{Prelovsek16, Kozarzewski18} while the charges remain localized—spin-charge separation is observed \cite{Zakrzewski18}. Analysis of different initial states reveals \cite{Protopopov19} that initial states with only doubly occupied or empty sites
seem to be many-body localized, indicating that the Hubbard model may host localized and delocalized states simultaneously. On the other hand, full many-body localization can be restored by breaking spin rotational symmetry, either by using uncorrelated disorder or different tunneling coefficients. This has also been confirmed in an independent study \cite{Sroda19}. Let us stress that these findings of ``full'' MBL should be understood in the weak sense (see the discussion of the thermodynamic limit above and in \cite{Sierant25}). Let us also note that
a detailed analysis allows for a partial construction of spin and charge LIOMs \cite{Thompson23} for the Fermi-Hubbard model. 

\subsection{Disordered Bose-Hubbard}

 Consider replacing fermions with bosons and investigate the on-site disordered Bose-Hubbard chain:
 \begin{equation}
\hat{H}
= - t\sum_{\left<i, j \right>}^{L} \hat{b}_{i}^{\dagger}\hat{b}_{j} 
 + \frac{U}{2} \sum_{i}^{L} \hat{n}_{i} (\hat{n}_{i}-1) + \sum_{i}^{L} h_{i} \hat{n}_{i}, \quad [\hat{b}_{i}, \hat{b}_{j}^{\dagger}]=\delta_{i,j}, \quad n_i = \hat{b}_{i}^{\dagger}\hat{b}_{i}.
\label{eq:bh}
\end{equation}
Its two-dimensional version was considered experimentally \cite{Choi16}, showing signs of localization on an intermediate time scale. An interesting small-scale experiment considered level spacing with interacting photons \cite{Roushan17}. Logarithmic growth of entanglement as well as the behavior of long-range correlations in small systems were also analyzed experimentally \cite{Lukin19, Rispoli19}. Theoretically, the disordered Bose-Hubbard model has been much less extensively studied than spinless fermions (i.e., the XXZ chain) for a simple reason: while for the latter, the local Hilbert space dimension is just two (four for the Fermi-Hubbard model), for bosons, the local limit is the total number of particles. Probably the most notable attempt is the description of bosonic systems in two dimensions using the approximate
tensor network approach \cite{Wahl19} (still restricting the model to
maximally double occupations of sites). Earlier treatment \cite{Sierant18}
considered both small and large systems in one dimension, with a much higher cutoff on the number of bosons per site. The study revealed
 the existence of a reverse mobility edge for 3/2 filling of the system: 
the higher-lying energy states were localized
for a lower amplitude of disorder for sufficiently strong
interactions. Many-body localization with superconducting
circuits, described by the Bose-Hubbard model with attractive interactions, was also discussed \cite{Orell19}. The
same system was also considered at half-filling \cite{Hopjan19, Yao20}, confirming the existence of a mobility edge. 
The recent study \cite{Chen24} stressed the fact that while the low-energy part of the spectrum for the half-filled Bose-Hubbard model shows a great similarity to the XXZ chain, the high-lying part of the spectrum is different, with separated clusters of bosons (due to high occupation of single sites) with a seemingly localized character. In this summary for bosons, let us also mention works on MBL with randomly interacting
bosons \cite{Sierant17,Sierant17b}, where no on-site disorder is necessary but $U$ in \eqref{eq:bh} is random. This work, similarly to the parallel effort for fermions \cite{BarLev16}, indicates that single-particle Anderson localization is not a necessary ingredient for the existence of MBL when interactions are turned on.

\section{Floquet models}

The existence of localization in periodically driven systems has been a standard highlight of single-particle quantum chaos, with the quantum rotator serving as a primary example; see, e.g., {\color{red} The Quantum Kicked Rotor: A Paradigm of Quantum Chaos} chapter \cite{Benenti26} in this volume. The seminal papers \cite{Fishman82, Grempel84} linked dynamical localization (in energy or momentum) to the Anderson localization problem in a disordered one-dimensional chain. For a generic many-body system, one expects, on general grounds, that periodic driving leads to unbounded heating of the system. It is, however, not the case. Already in the pre-MBL era, in a seminal contribution, T.~Prosen reported the transition between ergodic and integrable behavior in a periodically kicked chain of fermions \cite{Prosen98}. The interaction between fermions was included by standard nearest-neighbor interactions occurring, however, only at periodic kicks. 
A similar localization transition was later studied for different fully interacting models \cite{Dalessio13, Ponte15, Abanin15, Lazarides15, Abanin16} in finite-size systems. While in \cite{Prosen98, Ponte15} the control parameter that governed the crossover between phases was the strength of the kick, it could be translated into the frequency domain—the high frequency of the perturbation may result in a converging Magnus expansion and integrable effective dynamics, as discussed in \cite{Dalessio13} and generalized in \cite{Alessio14, Bukov15}. 

Floquet systems suffer, however, from similar problems as time-independent models when the thermodynamic limit is concerned. To see this, let us consider the kicked Ising model (KIM) \cite{Prosen02} defined by a unitary evolution operator over a single period:
\begin{eqnarray}
 U_{\mathrm{KIM}}= e^{-i g \sum_{j=1}^L \sigma^x_j} e^{ -i \sum_{j=1}^L (J \sigma^z_j \sigma^z_{j+1}  +  h_j \sigma^z_j ) }, \quad h_j\in [0,2\pi]
 \label{eq:KIM}
\end{eqnarray}
where $\sigma^{x,y,z}_j$ are Pauli operators and periodic boundary conditions are assumed. Let us assume $g=J=1/W$; $W$ plays the role of the disorder strength in the system.
The KIM is maximally ergodic for $W=4/\pi$ \cite{Akila16, Kos18}. For larger $W$ (smaller $g$), the second part of the evolution operator $U_{\mathrm{KIM}}$ dominates, leading to strong disorder and localization for finite systems. Using the state-of-the-art POLFED algorithm \cite{Sierant20p}, one can find eigenvectors $\ket{\psi_n}$ and the eigenvalues $e^{i \phi_n }$ of $U_{\mathrm{KIM}}$ up to system size $L=20$.

 \begin{figure}
 \includegraphics[width=0.99\linewidth]{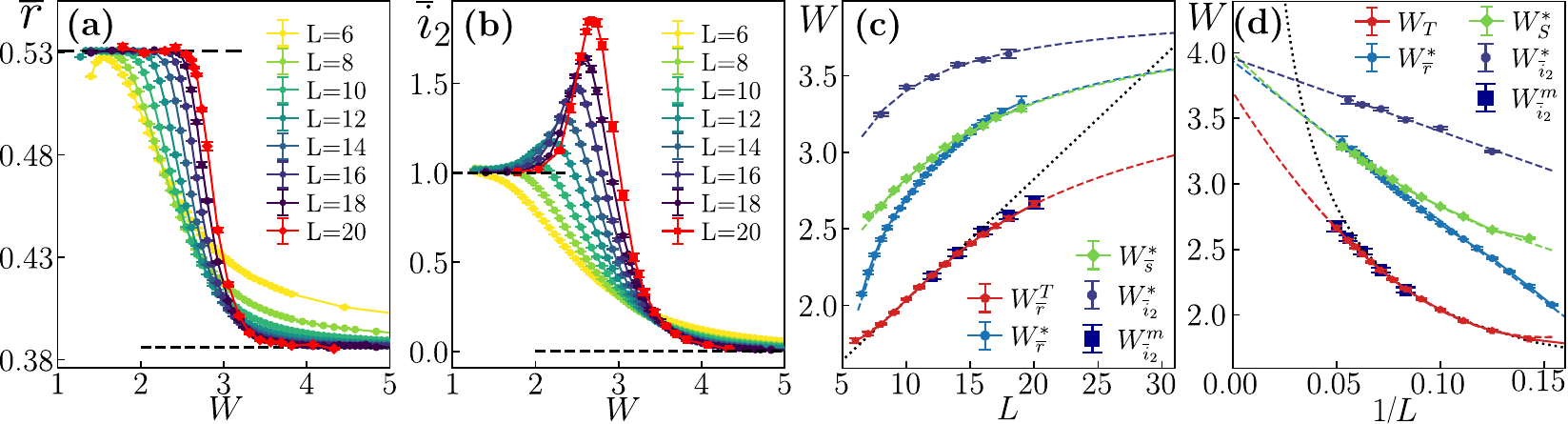} 
  \caption{The ergodic-MBL crossover in KIM \eqref{eq:KIM}. 
 Gap ratio $\overline r$ \textbf{(a)} and rescaled QMI $\overline{i}_2$ \textbf{(b)} as a function of disorder strength $W$ for different system sizes $L$; dashed lines mark ergodic and integrable predictions. Panels \textbf{(c)} and \textbf{(d)} show the disorder value $W^T_{\overline r}$ at which the mean gap ratio, $\overline r$, departs from the ergodic value
 as a function of $L$ and $1/L$, respectively. The same panels show the crossing points $W^*_X$ where $X$ stands either for $\overline r$, the rescaled entanglement entropy $ \overline s$, or the rescaled QMI $\overline{i}_2$. The dotted lines indicate $W_T(L) \sim L$ scaling; the dashed lines correspond to fits $W(L)=W_{\infty}+a/L+b/L^2$ with $W_{\infty}=3.97\pm0.03$ for $W^*_{\overline r}(L)$, $W^*_{\overline s}(L)$, and $W^*_{\overline{i}_2}(L)$. Figure reprinted from Sierant et al. \cite{Sierant22floquet}, copyright American Physical Society, 2023.
 }\label{figCOMB1}
\end{figure}   

The eigenvalues allow for the determination of the mean gap ratio $\overline r$.
As shown in Fig.~\ref{figCOMB1}, it displays a characteristic transition from ergodic to localized values, a transition dependent on the system size. In particular, $W^T_{\overline r}$, i.e., the disorder strength value at which the mean gap ratio departs from the ergodic asymptote, grows {\it linearly} with the system size $L$. On the other hand, extrapolating the crossings of different system size curves is convincingly done by a $W^*_{\overline r}(L)=W_{\infty}+a/L+b/L^2$ fit. Similar results are obtained for eigenvector characteristics such as the crossings of the $\overline i_2$, i.e., the rescaled quantum mutual information. The latter is defined as follows.
The entanglement entropy of eigenvectors $\ket{\psi_n}$ is given by $S(A)=-\sum_{i=1}^{i_M} \alpha_{i}^2 \log(\alpha_{i}^2)$, where $\alpha_{i+1}>\alpha_i$ are Schmidt basis coefficients \cite{Karol} of the eigenstate $\ket{\psi_n}$
for a partition of the 1D lattice into a subsystem $A$ and its complement.
Choosing $A=[1, L/2]$,  the rescaled entanglement entropy $\overline s = \left \langle S(A) \right \rangle/S_{COE}$ is calculated 
by taking the average $\left \langle . \right \rangle$ over the eigenstates and disorder realizations and rescaling the result by the average entanglement entropy $S_{COE}$ of eigenstates of the Circular Orthogonal Ensemble of random matrices (COE) that models the properties of $U_{\mathrm{KIM}}$ in the ergodic regime \cite{Alessio14}.
The quantum mutual information (QMI) is also depicted in Fig.~\ref{figCOMB1}. It is defined as  $I_2 = S(B)+S(C)-S(B\cup C)$ for subsystems $B=[1, \left \lceil{ L/4}\right \rceil ]$, $C=(2\left  \lceil{ L/4}\right\rceil, 2\left  \lceil{ L/4}\right\rceil+\left \lfloor{ L/4}\right \rfloor ]$ (where $\left  \lceil{ .}\right\rceil$, $\left \lfloor{ . }\right \rfloor$ denote the ceil and floor functions). The rescaled QMI is obtained as
$\overline i_2 = \left \langle I_2 \right \rangle / I_{COE}$, where $I_{COE}$ is the average QMI for COE eigenstates. 

\section{Quantum versus classical computing}

While signatures of MBL are observed in small systems, it is clear from the above that no evidence for MBL in the thermodynamic limit seems to exist. The mathematical proofs of its existence \cite{Imbrie16, Imbrie17} do not seem to be generally accepted. The state-of-the-art numerical studies (see, e.g., \cite{Sierant22challenges,Evers23Internal}) do not provide definite conclusions, as reviewed in \cite{Sierant25}. An obvious question is whether quantum computing could provide a necessary advantage in this direction. Some first answers may come from studies of nonstabilizerness. Nonstabilizerness (also known as quantum magic) is a quantum resource enabling the quantification of the complexity of quantum states. So-called stabilizer states \cite{Gottesman98} can be efficiently simulated on classical computers, despite sometimes being highly entangled. Quantum magic quantifies how far a given state is from the set of stabilizer states. In the dynamics of ergodic systems, initial information is soon lost; the wavefunction resembles a random vector. Interestingly, for the present review, the behavior of magic in finite systems exhibiting MBL has also been addressed \cite{Falcao25},
showing strong correlations between the growth of magic and entanglement.

Magic is analyzed using stabilizer Rényi entropy (SRE) \cite{Leone22}, defined as
\begin{equation} \label{eq:sre1}
\mathcal{M}_k(|\Psi\rangle) = \frac{1}{1-k}\log_2\left[\sum_{P\in \mathcal{P}_{L}}\frac{\langle \Psi|P|\Psi \rangle^{2k}}{D}  \right],
\end{equation}
where $L$ is the number of qubits, $k$ is the Rényi index, and $P$ is a Pauli string that belongs to the Pauli group $\mathcal{P}_L$. We consider here $\mathcal{M}_2$ only. 
For random unitary circuits, the SRE saturates to the Haar-random state value~\cite{Turkeshi23a}:
 $   \mathcal{M}_2^{\mathrm{Haar}} = \log_2 (D + 3) - 2$,
where $D=2^L$ is the Hilbert space dimension. 

 \begin{figure}
 \includegraphics[width=0.45\linewidth]{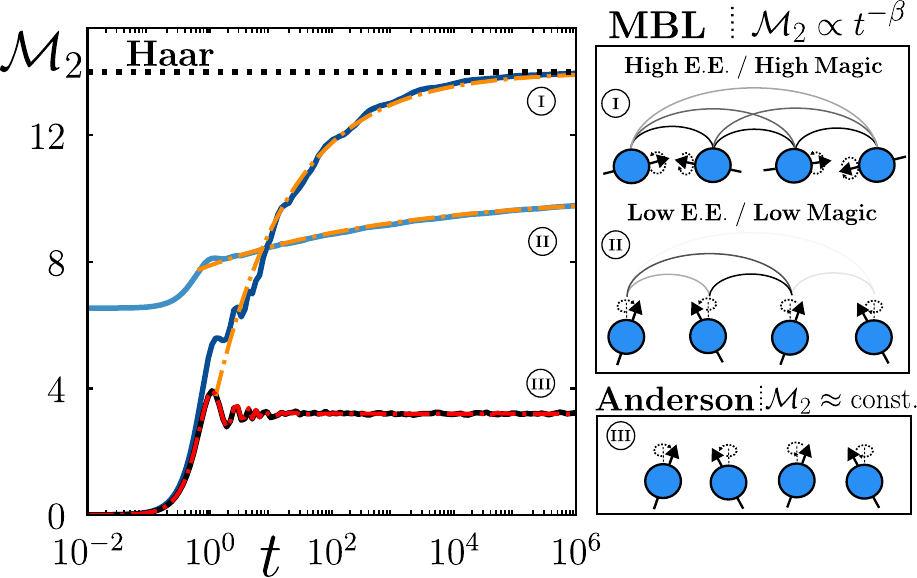} 
 \includegraphics[width=0.54\linewidth]{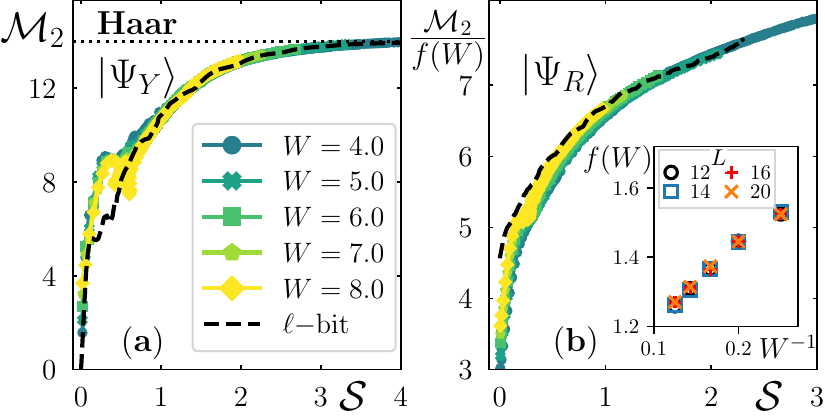} 
  \caption{ The left panel visualizes the correlation between entanglement and magic as measured by SRE in the MBL regime, as obtained from the dynamics of the l-bit model~\eqref{eq:l-bit} for different states. Case (III) corresponds to a non-interacting Anderson insulator. The right panel shows that the 
 correlation between $\mathcal{M}_2$ and $S$ also exists for the TFIM in the MBL regime. For further explanations, see the text.
 Figure reprinted and modified from Falcao et al. \cite{Falcao25}, copyright American Physical Society, 2025.
 }\label{fig:mag}
\end{figure}   

We consider first the dynamics of genuine MBL as realized by the l-bit model~\eqref{eq:l-bit}. Setting $J_{ij..}=0$, we realize the Anderson insulator. Then both entanglement entropy, $S$, and $\mathcal{M}_2$ saturate fast at low values (the latter may even be found analytically \cite{Falcao25}). For the interacting case, saturation is also observed, but it occurs at a slower pace and follows the power-law behavior $   \mathcal{M}_2^{\mathrm{MBL}} = \mathcal{M}_2^{\mathrm{sat}} - c/t^{\beta}$, with $\mathcal{M}_2^{\mathrm{sat}}$, $c$, and $\beta$ depending on the initial state. For example, an $X$-polarized state $|\Psi_X^{+}\rangle = \bigotimes_{k=1}^{L} (|\downarrow\rangle + |\uparrow\rangle)/\sqrt{2}$
leads to a rapid growth of $\mathcal{M}_2$ up to the Haar value, while a simple separable state of up and down spins $\tau^i$ in \eqref{eq:l-bit} results in much slower growth (case II in Fig.~\ref{fig:mag}).

Similar trends are observed not only for the l-bit model but also in a microscopic model, the well-known disordered transverse-field Ising model (TFIM). As long as the random disorder is sufficiently large to put our finite system in the MBL regime, there is a simple direct correspondence between the entanglement entropy and SRE, as shown for the $|\Psi_Y\rangle$ state (polarized along Y) (middle panel in Fig.~\ref{fig:mag}) or $|\Psi_R\rangle$ (right panel in Fig.~\ref{fig:mag}). The latter state is constructed by combining random rotations on Bloch spheres and has $\mathcal{M}_2>0$ for $t=0$, as it cannot be obtained by Clifford gates.
Here, to get the collapse observed in the figure, an additional rescaling factor $f(W)$, independent of system size, was needed.

The strong correlations between entanglement entropy and SRE in the MBL regime suggest that in the MBL regime either $S$ or $\mathcal{M}_2$ may serve as an intrinsic evolution time (see Sec.~\ref{intri}). Thus, to reach long times (for large sizes) necessary to validate the existence of MBL in the thermodynamic limit, one needs to deal with states of significant magic, most possibly requiring quantum computers.

\section{Quantum scars are not MBL}

While discussing localization in interacting many-body systems, one must address the notion of quantum many-body scars, which has received significant interest recently \cite{Bernien17}. Let us recall that here, for a specific initial state, one could observe a characteristic decay of correlations with a pronounced damped oscillatory character instead of a simple damping. The possible origin of such unexpected behavior is the fact that the initial state has a large overlap with a family of regular, roughly equally spaced in energy, states that are only weakly coupled to the rest of the system, which otherwise seems ergodic. While several model studies of such behavior have appeared, let us mention here the pioneering work \cite{Shiraishi17} as well as a review \cite{Moudgalya22rev} on the topic. Quantum many-body scars represent situations where regular, ``many-body scar'' states are embedded in an ergodic spectrum and are, therefore, intrinsically different from MBL, where all states are supposed to be localized. Therefore, we do not discuss quantum scars here, referring the interested reader to \cite{Moudgalya22rev,Chandran23rev}. 

\section{Conclusions}
While the existence of MBL in the thermodynamic limit cannot be considered proven beyond a reasonable doubt, real laboratory systems are finite. For typical systems of a few tens of spins, MBL has been shown to exist, forming arguably the most robust example of ergodicity breaking in interacting many-body systems. While it was studied primarily with disorder, it has also been shown to exist in its absence, for example, in tilted spin chains (lattices). While interesting on its own, it may have practical implications for preserving memory in quantum devices or reducing heating in driven systems. Its application to facilitate quantum computing has also been suggested \cite{Wang22}. While much is known about this phenomenon (and only a partial account is presented here), it will continue to yield novel, surprising ideas and problems for study.

\begin{ack}[Acknowledgments]

The author acknowledges long-term collaboration on various problems, including those reviewed in this work, with the late Dominique Delande.
He also thanks Monika Aidelsburger, Luca Barbiero, Titas Chanda, Marcelo Dalmonte, Nicolas Laflorencie, Maciej Lewenstein, Mateusz \L{}\c{a}cki, Marcin Mierzejewski, Giovanna Morigi, Krzysztof Sacha, Luis Santos, Antonello Scardicchio, Luca Tagliacozzo, Emanuele Tirrito, Lev Vidmar, and Ruixiao Yao for discussions and common work. The latter would not be possible without my former and present students suffering the interactions with me on this topic - Adith S. Aramthottil, Pedro R.N. Falca\~o, Konrad Pawlik, Maksym Prodius, Tomasz Szołdra. Special thanks go to a long-time student and later a leading collaborator, Piotr Sierant. This work was funded by the National Science Centre, Poland, under the OPUS call within the WEAVE program 2021/43/I/ST3/01142.
\end{ack}

\bibliographystyle{JHEP}%

\providecommand{\href}[2]{#2}\begingroup\raggedright\endgroup

\end{document}